\documentclass[pldi-cameraready]{sigplanconf-pldi16}
\usepackage{SIunits}            
\usepackage{courier}            
\usepackage[scaled]{helvet} 
\usepackage{url}                  
\usepackage{listings}          
\usepackage{enumitem}      
\usepackage[colorlinks=true,allcolors=blue,breaklinks,draft=false]{hyperref}   
\newcommand{\doi}[1]{doi:~\href{http://dx.doi.org/#1}{\Hurl{#1}}}   

\usepackage{textcomp}
\usepackage{graphicx}
\usepackage{booktabs}
\usepackage{titlesec}

\newcommand{\given}{\mid}

\titlespacing*{\section}{0pt}{9pt plus 0pt minus 3pt}{6pt plus 0pt minus 3pt}
\titlespacing*{\subsection}{0pt}{9pt plus 0pt minus 3pt}{6pt plus 0pt minus 3pt}
\titlespacing*{\subsection}{0pt}{9pt plus 0pt minus 3pt}{6pt plus 0pt minus 3pt}

\begin{document}

\title{SIMPL: A DSL for Automatic Specialization of Inference Algorithms}

%
%

\authorinfo{Rohin Shah\textsuperscript{1} \and Emina Torlak\textsuperscript{2} \and Rastislav Bod\'{i}k\textsuperscript{2}}
{\textsuperscript{1}University of California, Berkeley \and \textsuperscript{2}University of Washington}
{rohinmshah@berkeley.edu \and emina@cs.washington.edu \and bodik@cs.washington.edu}

\maketitle

\begin{abstract}
Inference algorithms in probabilistic programming languages (PPLs) can be thought of as interpreters, since an inference algorithm traverses a model given evidence to answer a query.  As with interpreters, we can improve the efficiency of inference algorithms by compiling them once the model, evidence and query are known.  We present SIMPL, a domain specific language for inference algorithms, which uses this idea in order to automatically specialize annotated inference algorithms.  Due to the approach of specialization, unlike a traditional compiler, with SIMPL new inference algorithms can be added easily, and still be optimized using domain-specific information.  We evaluate SIMPL and show that partial evaluation gives a 2-6x speedup, caching provides an additional 1-1.5x speedup, and generating C code yields an additional 13-20x speedup, for an overall speedup of 30-150x for several inference algorithms and models.
\end{abstract}

\section{Introduction}

Probabilistic programming languages (PPLs) extend regular programming languages with random variables and the ability to observe evidence about these random variables.  Thus, a probabilistic program specifies a probability distribution over random variables.  The programmer can then ask queries about the random variables, which the PPL answers using \textit{inference algorithms}.  The goal of probabilistic programming is to make probabilistic modeling and machine learning accessible to programmers without expertise in these areas.  As a result, PPLs must hide the inference algorithm from the programmer, and so the inference algorithms must be general enough to work with all models that programmers could write.  Many such PPLs have been created along these lines, such as BLOG \cite{BLOG}, Church \cite{Church}, Figaro \cite{Figaro} and Stan \cite{Stan}.

However, the generality of PPLs comes at a price---the inference algorithms are orders of magnitude slower than custom inference algorithms written for specific problems.  Many PPLs are implemented as interpreters and are not compiled, and so suffer from additional overhead.  Specializing these inference algorithms to the particular model and query of interest can thus lead to large speedups.  Even for simple models, where the abstractions are simple and low-cost, this specialization can produce speedups over a hand coded generic implementation in C.

There are many different inference algorithms, and new ones are created frequently.  Each inference algorithm is useful for a certain class of models, evidence and queries, and so many PPLs implement several inference algorithms.  For example, BLOG \cite{BLOG} implements rejection sampling, likelihood weighting, MH, MCMC, and particle filtering, Stan \cite{Stan} implements Hamiltonian Monte Carlo and No-U-Turn Sampling, and Figaro \cite{Figaro} implements variable elimination, belief propagation, importance sampling, MH, particle filtering, parameter learning through expectation maximization, and more.  In addition, there are classes of inference algorithms that use extra information to scale up to bigger models or bigger data, such as lifted inference, variational inference, structured inference, and combinations of these.  Researchers also create new proposal distributions in order to get better results, and these must be optimized as well.

Thus, there are two requirements for PPLs to become competitive with hand-coded solutions---first, inference algorithms must be optimized to run orders of magnitude faster, and second, it needs to be possible to easily add new optimized inference algorithms to a PPL.  We propose to solve this problem by \textit{partially evaluating} a na\"{\i}vely written inference algorithm in order to obtain fast compiled programs.  Specifically, we make the following contributions:

\begin{enumerate}
\item We determine that important optimizations for inference algorithms are partial evaluation (PE), caching, and data structure choice.
\item We note that creating an optimized, \textit{self-specializing} model data structure enables much better PE.
\item In order for PE to be effective, we need to be able to unroll loops to expose more optimization opportunities.  However, unrolling loops with many iterations leads to code blowup.  We identify a way to get the benefits of unrolling while avoiding code blowup---ensure that the model structure will be small and unroll only loops over only the model structure.
\item We present SIMPL: Scalable Inference for Modern Probabilistic Languages, a domain specific language that uses these ideas to optimize inference algorithms.  Algorithms written in SIMPL are \textit{simple}, that is, they are similar to their textbook presentations.  SIMPL applies PE and caching, guided by programmer annotations, to generate \textit{fast} C code, comparable to hand-optimized code.
\end{enumerate}

\section{Overview} \label{sec:overview}

Evaluation of a probabilistic program can be thought of as a two-stage process.  In the first stage, the probabilistic program is compiled into an intermediate representation called a \textit{probabilistic model}.  In the second stage, an inference algorithm is run on the model in order to answer the query.  We focus only on the second stage, as that is the bottleneck.

\subsection{Current Best Practice}

Currently, there are two main ways in which generic inference algorithms are written.  They can be written in the obvious, na\"{\i}ve way, with little or no attention to performance, which results in a lot of interpretation overhead, as in the BLOG interpreter and the original Church implementation.  On the other hand, they can be written as \textit{self-specializing algorithms}, that is, algorithms that generate compiled code, which when run produces the answer.  This approach eliminates the interpretation overhead, leading to fast code, but it takes a lot of implementation effort.  This approach can be seen in Swift (the BLOG compiler) and Bher (a compiler for Church) \cite{Bher}.  However, Swift only supports two inference algorithms, and Bher only supports one.  It requires too much implementation effort to create self-specializing algorithms, especially when there are so many inference algorithms that need to be implemented.

SIMPL introduces a third approach---the algorithm writer writes the algorithm in the obvious way, and it is automatically converted into a self-specializing algorithm by SIMPL.  This give us the best of both approaches---we get the speed of self-specializing algorithms as well as the ease of implementation of na\"{\i}ve algorithms.

\subsection{Key Ideas}

Key ideas behind SIMPL are the identification of optimizations that are relevant for inference, in particular that PE can be used to "compile" inference algorithms, and the optimization of the model data structure to decrease annotation burden on the inference algorithm writer.

\subsubsection{Partial Evaluation}

Since na\"{\i}ve inference algorithms are like interpreters, we can use the first Futamura projection \cite{Futamura} to get a self-specializing algorithm that produces compiled code.

Specifically, an inference algorithm will typically take many samples, and in taking each sample it will repeat work, such as traversing the model, looking up the parents of a node, indexing into a conditional probability table, etc.  If we think of the inference algorithm as an interpreter, then we can think of the model, evidence and query as the program that is being interpreted.  Thus, we can achieve a self-specializing algorithm by \textit{partially evaluating} the inference algorithm with respect to the model, evidence and query.  This allows us to generate optimized code for generating samples.  Since we generate thousands or even millions of samples, PE can save a lot of work, leading to substantial speedup.

However, PE must make decisions about what to evaluate and what to leave until runtime.  This is typically done through binding-time analysis and a conservative strategy for inlining function calls and unrolling loops.  These program analyses and heuristics are often suboptimal.  So, in SIMPL, we use programmer annotations to instead guide the optimization.  We expect that the additional annotation effort is worth the gain in performance, because fast inference algorithms are crucial for a PPL.

\subsubsection{Model Data Structure}

Many inference algorithms operate on the same model structure, and so it is worthwhile to put extra effort into the model data structure for a corresponding decrease in effort for inference algorithms.  In particular, the model is already annotated with information that can be useful for inference algorithm optimization---for example, SIMPL knows that the values of evidence nodes are fixed and do not change.  This reduces the annotation burden on the programmer, who no longer has to specify that the evidence is known at compile time.  The model is also \textit{self-specializing}---it can generate a faster version of itself at runtime that supports only the necessary operations.

In addition, it is important that the model data structure is linear in the size of the probabilistic program that generated it.  This allows us to be certain that the model will be of a reasonable size, and to unroll loops over the model without worrying about code blowup.

\subsubsection{Relevant Optimizations}

Through analysis of a self-specializing algorithm for Gibbs sampling, we found that the crucial optimizations were PE, caching, good data structure choice, and elimination of high-level language overhead (such as infinite precision arithmetic).  SIMPL provides PE and caching through annotations.  Good data structures for inference algorithms are generated by the self-specializing model data structure.  High-level language overhead is eliminated by generating C code.

\subsection{Example: Likelihood Weighting} \label{sec:likelihood}

We illustrate SIMPL by implementing the likelihood weighting algorithm for Bayes Nets where each node takes on a boolean value, and show that by adding a few annotations, SIMPL will optimize the algorithm to get large speedups.

\subsubsection{Bayes Nets}

\begin{figure}
  \centering
  \includegraphics[scale=0.5]{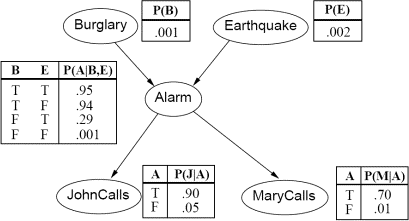}
  \caption{Bayes Net for the Burglary model.}
  \label{fig:burglary}
\end{figure}

Some probabilistic programs can be compiled into \textit{Bayes Nets}, after which inference can be run.  A Bayes Net represents a set of random variables and their dependencies using a Directed Acyclic Graph (DAG).  For example, consider the Burglary model, illustrated in Figure~\ref{fig:burglary}.  An Alarm could be triggered by both a Burglary and an Earthquake, which may result in the neighbors John and Mary calling.  We denote these variables by $A$, $B$, $E$, $J$ and $M$ respectively.  Every node is labelled with a \textit{conditional probability table} (CPT) that specifies how that node depends on its parents.  For example, if a Burglary is in progress but there is no Earthquake, the probability of the Alarm ringing can be found from the Alarm's CPT:  $P(A \given B, \neg E) = 0.94$.  Given this model, we can ask \textit{queries}, such as the query $P(B \given J, M)$.  In this query, the \textit{evidence} is that both John and Mary have called, and the query variable is Burglary.  Such a query can be answered by an inference algorithm, such as likelihood weighting.

\subsubsection{Using Conditional Probability Tables}

\begin{figure}
\begin{lstlisting}
; Returns P(node | parents(node))
(define (cp node)
  ; Calculate P(node = true | parents(node))
  (let ([base (true-cp node)])
    ; P(node = false) = 1 - P(node = true)
    (if (value node) base (- 1 base))))

; Returns P(node = true | parents(node))
(define (true-cp node)
  (index (CPT node) (parents node)))

; Indexes into a CPT based on parents' values
(define (index cpt parents)
  (if (null? parents)
      cpt
      (if (value (car parents))
          (index (first cpt) (cdr parents))
          (index (second cpt) (cdr parents)))))

; Example: P(A | B, not E)
> (list (value Burglary) (value Earthquake))
'(#t #f)
> (index '((0.95 0.94) (0.29 0.001))
         (list Burglary Earthquake))
0.94
\end{lstlisting}
\caption{Code to index into a conditional probability table.}
\label{fig:cp}
\end{figure}

We represent the Bayes Net as a graph where each node has a boolean value, a CPT, a list of parent nodes, and a list of children nodes.  The CPT is represented as a nested list structure, for example, the CPT for Alarm would be represented by the nested list \textquotesingle\texttt{((0.95 0.94) (0.29 0.001))}.  We first write a procedure \texttt{true-cp} which computes the probability of a node given its parents values by picking out the correct number from the CPT.  The code is shown in Figure~\ref{fig:cp}---it simply traverses the nested list structure recursively, using the value of the parent node to choose which element of the list to consider.

In likelihood weighting, we will often call \texttt{true-cp} on the same node, but with the parents having different values.  For example, we sometimes want to know $P(A \given B, \neg E)$, and sometimes $P(A \given \neg B, E)$.  If we simply run the procedure above twice, a lot of work will be repeated, such as looking up the parents and the CPT of Alarm, and the recursing through the list of parents, and indexing into the CPT of Alarm.  Instead, we can \textit{partially evaluate} the \texttt{true-cp} procedure on \texttt{Alarm}, assuming that we do not know the values of the parent nodes yet.  This yields code that, when given the values of Burglary and Earthquake, returns the entry in Alarm's CPT corresponding to those values.

SIMPL assumes by default that in inference algorithms, the values of nodes that are not evidence will change but the structure of the graph will not, and so it can perform this optimization automatically without any annotations.  These assumptions can be overridden by the algorithm writer as explained in Section 3, but they work well for this code.  SIMPL will generate the code shown in Figure~\ref{fig:likelihood-gen} (see the box marked \lstinline{(true-cp Alarm)}).  Notice that in the generated code, the graph structure has been eliminated, the nested list of the CPT is gone, and there is no more recursion.

\subsubsection{Generating Samples}

\begin{figure}
\begin{lstlisting}
; Samples non-evidence nodes in the Bayes net.
(define (sample bayesnet evidence)
  ; For every node
  (for/unroll ([node bayesnet])
    ; If it is not evidence
    (unless (member node evidence)
      ; Resample the node
      ; (flip p) gives #t with probability p
      (set-value! node (flip (true-cp node))))))

; Calculates the weight of a sample.
(define (weight evidence)
  ; Multiply probabilities of all the evidence
  (foldl * 1 (map cp evidence)))
\end{lstlisting}
\caption{Weighted sample generation.}
\label{fig:sample}
\end{figure}

Now that we can manipulate CPTs, we can generate samples.  The likelihood weighting algorithm generates weighted samples by processing nodes in topological order, and sampling values for nodes according to the probabilities in their CPTs.  Evidence nodes are not sampled (since their values are known), but their probabilities are incorporated to calculate the weight.  This can be done with the code in Figure~\ref{fig:sample}.  The code has the annotation \lstinline{for/unroll}, which tells SIMPL to unroll the for loop at PE time instead of leaving it until runtime (the default).  Unrolling the loop exposes more opportunities for PE (since the loop variable is a concrete value), but it can lead to code blowup.  In this case, since the loop is unrolled, SIMPL is able to determine whether or not a node is evidence, allowing it to eliminate the branch on \lstinline{(member node evidence)}.  It is also able to partially evaluate the call to \lstinline{true-cp} since the argument is concrete.  SIMPL generates the code in the box marked \lstinline{(sample burglary-model evidence)} in Figure~\ref{fig:likelihood-gen} for taking a sample, and generates the code in the box marked \lstinline{(weight evidence)} for calculating the weight of a sample.

Note that if we used \lstinline{for} instead of \lstinline{for/unroll}, SIMPL would not be able to perform any specialization, since the value of the \lstinline{node} variable would not be known at PE time.

\begin{figure}
\begin{lstlisting}
; Answers a query by taking weighted samples.
(define (likelihood query evidence net N)
  ; Array storing total weight of the samples
  (let ([weights (vector 0 0)])
    ; Repeat N times
    (for ([j N])
      (sample net evidence)
      ; Choose the bin based on the query
      (let ([idx (if (value query) 0 1)])
        ; Add the weight of the current sample
        (vector-set! weights idx
                     (+ (vector-ref weights idx)
                        (cache
                         (weight evidence))))))
    ; Normalize so the probabilities add to 1.
    (normalize weights)))
\end{lstlisting}
\caption{Likelihood weighting algorithm.}
\label{fig:likelihood}
\end{figure}

\begin{figure*}
\centering
\includegraphics[scale=0.45]{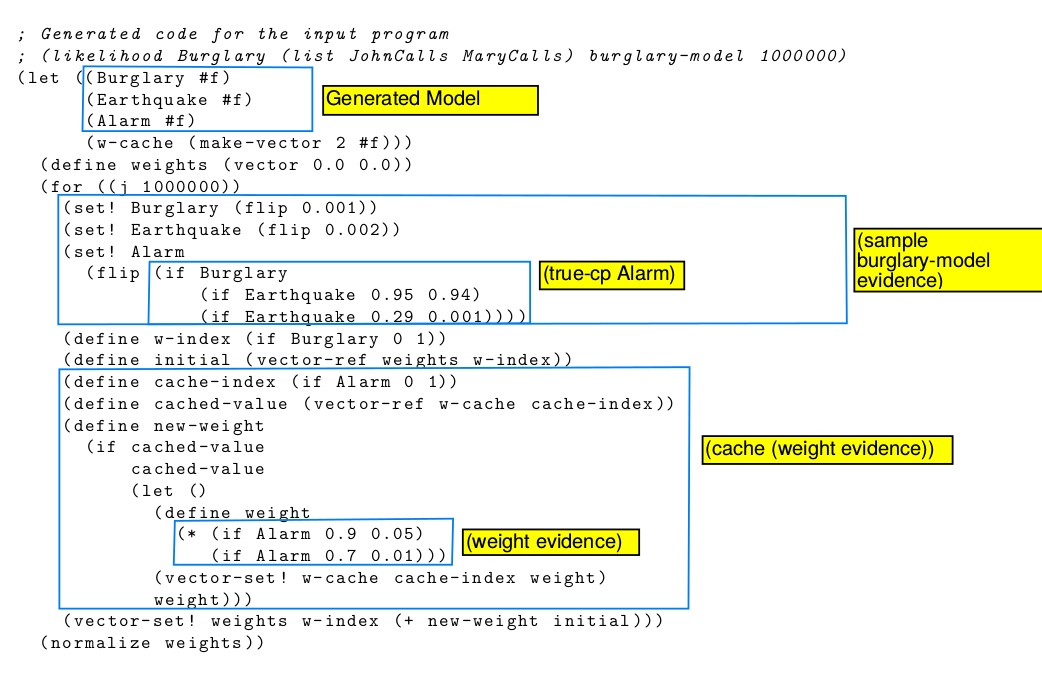}
\caption{Code generated when Figure~\ref{fig:likelihood} is specialized with respect to the query $P(B \given J, M)$.  The generated code has been simplified manually in order to aid readability (renaming variables, eliminating temporary variables, and eliminating superfluous lets).  Portions of the code have been annotated with the expressions that generated them.}
\label{fig:likelihood-gen}
\end{figure*}

Finally, we can take a lot of samples and use them to answer the query, as shown in the \lstinline{likelihood} function in Figure~\ref{fig:likelihood}.  The generated code for the query $P(B \given J, M)$ is shown in Figure~\ref{fig:likelihood-gen}.  Note that we use a normal \lstinline{for} loop instead of \lstinline{for/unroll}, and as a result the generated code still has the same \lstinline{for} loop.  (If we tried to unroll the \lstinline{for} loop, we would get a huge code blowup.)

In addition, we use the \lstinline{cache} annotation in order to remember previous values of the weights of samples.  Like the \lstinline{memo} procedure in other languages, the \lstinline{cache} annotation remembers previous values of an expression in a cache.  For the Burglary model, SIMPL will determine that the weight is given by $P(J \given A) * P(M \given A)$.  Since $J$ and $M$ are evidence, the weight depends only on the value of $A$.  So, the cache only ever contains two values, one for each possible value of \lstinline{Alarm}.

For this specific model and query, the caching is actually not beneficial, because the generated code for weight calculation is so simple and fast that the overhead of a cache is not worth it.  However, with larger models caching does provide benefits.  It would be interesting to allow a programmer to specify a model-dependent \textit{policy} which allows the annotations to vary with the model to account for cases like these, but we do not explore policies in this paper and leave it to future work.

\section{Architecture of SIMPL}

SIMPL consists of three layers of programs---the SIMPL core, written by us; inference algorithms, written by the implementers of PPLs, and the probabilistic program, written by the end user.

\subsection{SIMPL Core}

The SIMPL core implements the various optimizations, such as PE, caching, and data structure choice.  In particular, it also specifies the data structures used to represent the model, which can be thought of as the intermediate representation of the probabilistic program.

\subsection{Inference Algorithms}

Inference algorithms are written by the implementer of a PPL that is written in SIMPL.  An inference algorithm must take in a model, evidence, and query, and perform probabilistic inference to answer the query.  These algorithms are written in a simple, algorithmic fashion, along with a few annotations.  When written in this na\"{\i}ve way, these programs are usually short, and so the burden of annotation is not too large.  In addition, as part of the implementation of a PPL, the inference algorithms will be used many times, and so the one-time cost of annotation is amortized over many runs of the algorithm.

\subsection{Probabilistic Program}

The probabilistic program defines a particular probabilistic model and is written by the end user.  This is compiled into an instance of the model data structure in the SIMPL core, and then an inference algorithm is run to answer the query.  There are no annotations in the probabilistic program---the optimizations in SIMPL are completely hidden from the end user.

Our models are most similar to those of BLOG.  A BLOG program declares types of random variables through declaration statements.  During execution, there can be many instances of any given type.  According to the BLOG team, BLOG programs typically contain around 20 declaration statements, even those solving real-world problems, such as the model for NET-VISA \cite{Netvisa} which is used for international seismic monitoring.  For example, in the MultiBurglary model, there are three types of random variables---Burglary, Earthquake, and Alarm, and so the BLOG program describing it is very small.  There are a thousand possible houses, and so at runtime there are a thousand Burglary instances and a thousand Alarm instances.  Thus, in typical models, the \emph{structure} of the model is small, even though the data may be large.

In SIMPL, we ensure that the structure of the probabilistic graphical model is linear in the size of the corresponding BLOG program.  Then, for typical programs, this structure will be small.  This is crucial in order for PE to scale---it allows us to unroll loops over nodes without fear of code blowup, which exposes many additional optimization opportunities for PE.  In the likelihood weighting example of section \ref{sec:likelihood}, the two loops that were unrolled were precisely the two loops over the nodes in the model, and the recursive functions that were inlined were functions that recursed over the parents of a node, which are a subset of the nodes in the model.

\section{Partial Evaluation}




Partial evaluation (PE) is a technique that allows an input program to be \textit{specialized} to statically known inputs, to get a faster program that takes in the inputs known at runtime.  Formally, given a program with two inputs $P(x, y)$, where $x$ is fixed to a value $c$ and $y$ changes, PE generates a \textit{residual program} $P[x=c]$, such that
$$P[x=c](y) = P(c, y)$$
and $P[x=c]$ is faster than $P$.  If we run $P[x=c]$ for many different values of $y$, then we can achieve huge speedups.

Inference algorithms typically take a large number of samples and use these samples to answer a query.  The only difference between any two samples are the random numbers used in generating the samples.  Thus, the appropriate PE is to treat the sampling procedure at the program $P$, the random numbers as the unknown input $y$, and all other inputs (model, evidence, query) as the known inputs $x$.

PE has two main components---binding-time analysis and reduction.  Binding-time analysis labels each expression as either \textit{static}, meaning that it can be evaluated at PE time, or \textit{dynamic}, meaning that residual code will need to be generated.  Reduction uses the knowledge from the binding time analysis to compute statically known values and to generate the residual program.  SIMPL combines both stages together, called \textit{online} PE, because it produces faster residual code.  A detailed discussion of PE can be found in \cite{Sestoft}.

Common issues with PE are conservative binding-time analysis, code duplication, mutability, computation duplication, and non-termination \cite{Sestoft}.  SIMPL provides annotations that allow the programmer to control the PE, which can be used to avoid these problems.

\subsection{Annotations}

SIMPL supports annotations for controlling binding-time analysis---\lstinline{static} and \lstinline{lift}---as well as annotations for unrolling and inlining---\lstinline{for/unroll} and a modified version of \lstinline{define}.

In SIMPL, the default binding-time analysis is to make effectful computations dynamic (notably random number generation, mutable data and non-evidence node values), and to make all other atoms static.  Compound expressions are then treated as static or dynamic based on their subexpressions.  Since SIMPL performs online PE, the static/dynamic distinction is \textit{not} syntactic---it is possible for the same expression to be marked as static once and dynamic another time in a single program run.  For instance, in the \lstinline{true-cp} procedure, the expression \lstinline{(value (car parents))} is static if the first parent is evidence, and is dynamic otherwise, both of which can occur in a single program run.

Binding-time analysis can be overridden by the annotation \lstinline{(static expr)}, which tells SIMPL that \lstinline{expr} should be performed at PE time.  For example, if the programmer knows that the data in a particular array can be computed at PE time, (s)he can wrap all expressions involving that array with \lstinline{static} to cause it to occur at PE time.  Binding-time analysis can also be overridden by the annotation \lstinline{(lift expr)}, which causes the binding time of \lstinline{expr} to be dynamic.  Note that any subexpressions of \lstinline{expr} will retain their original binding time.  For example, in \lstinline{(+ 2 3)}, if we want to delay the addition until runtime, we would use \lstinline{(+ (lift 2) 3)}, \textit{not} \lstinline{(lift (+ 2 3))}.

Unrolling and inlining are controlled by \lstinline{for/unroll} and \lstinline{define}.  \lstinline{for/unroll} works just like a normal \lstinline{for} loop, but the loop iteration is performed at PE time, instead of at runtime.  An example was given in the \lstinline{sample} procedure in Figure~\ref{fig:sample}, and the unrolling of the loop at PE time allowed additional specialization of the body of the loop since the value of the iteration variable was known.  This annotation can lead to code blowup, because the body of the loop is evaluated for each value of the iteration variable, which could lead to the body being replicated multiple times in the generated code.

The default behavior of \lstinline{define} is to always inline the function.  This can also lead to code blowup and non-termination in the case of recursive functions, but we find that these are rare cases for inference algorithms.  For these rare cases where inlining is undesirable, the programmer can annotate procedure definitions to specify that a function call should not be inlined if particular arguments are symbolic.  This is particularly necessary for recursive functions that may be called with symbolic arguments---if inlining is not prevented, then SIMPL will go into an infinite loop, attempting to inline the recursive call again and again.  This was not necessary for the \lstinline{index} helper of \lstinline{true-cp} because none of the arguments to \lstinline{true-cp} were ever symbolic and so unwinding the recursion always eventually reached the base case, and so there were no infinite loops.

\subsection{Avoiding Problems of PE}

The use of annotations allows the programmer to avoid many of the common problems of PE.

\subsubsection{Conservative Binding-Time Analysis}

Typically, a binding-time analysis is conservative, erring on the side of marking expressions as dynamic, in order to be sound.  However, this means that less reduction can happen, and so the residual programs are not as fast as they could be.  SIMPL avoids the issue by using a default, conservative binding-time analysis, and then letting the programmer customize the defaults.  \lstinline{static} in particular forces an expression to be evaluated at PE time, and thus can fix an overly conservative binding-time analysis.

\subsubsection{Code Blowup}

Code duplication can occur either by unrolling a loop or inlining a function call (in both cases leading to duplication of the body).  This makes it easy for the programmer to avoid code duplication---simply never unroll any loops and never inline function calls.  However, unrolling and inlining expose more opportunities for PE, which can lead to large speedups.  There is thus a tradeoff between increased PE and code blowup.  So, we provide the programmer with annotations that can control the level of unrolling and inlining, which the programmer can use to find the right tradeoff.  A good heuristic is to only unroll loops over the model structure, because we are guaranteed that it will be of reasonable size.

\subsubsection{Mutability}

Since PE with mutability is difficult and is often very conservative, SIMPL never partially evaluates any reads from or writes to mutable data.  Instead, if this PE is needed, it can be achieved using \lstinline{static}.

\subsubsection{Computation Duplication}

Even for functional computations, PE with inlining can often lead to computation duplication, where an expression is inlined in a function call, resulting in the expression being duplicated and the computation being performed twice.  While this would not affect correctness, it does slow down the program.  This is fixed by introducing temporary variables to store results, as described in section \ref{sec:pe-prim-impl}.

\subsection{Correctness and Debugging} \label{sec:pe-soundness}

PE in SIMPL can go wrong in two ways---non-termination and bad binding-time analysis.

\subsubsection{Non-termination}

When SIMPL tries to inline a recursive or mutually recursive function with a symbolic argument, it will continually inline recursive calls and never reach the base case since the argument is symbolic.  Detection of such non-termination is undecidable in general, but in practice this should not be a problem.  Programmers will first develop and test their algorithms without optimizations.  If they then get an infinite loop when enabling PE, they can examine the stack trace and find the function that was being inlined infinitely.  Annotating the definition of that function would then fix the problem.

\subsubsection{Bad Binding-time Analysis}

A bad binding-time analysis occurs when the algorithm writer marks an expression as static, but it's value cannot be found at PE time.  For example, consider the following program:

\begin{lstlisting}
(for ([i 10])
  (print (static (+ i 1))))
\end{lstlisting}

Here, since the loop is not unrolled, the value of \lstinline{i} is unknown.  Thus, the \lstinline{(+ i 1)} operation cannot be fully evaluated at PE time, and so SIMPL raises an error.  In general, SIMPL will always raise an error when a \lstinline{static} is used incorrectly.  The inference algorithm writer can then fix the annotations as required.  For example, in the above program (s)he could either choose to unroll the loop, or (s)he could remove \lstinline{static} so that \lstinline{(+ i 1)} is allowed to emit residual code.

\subsection{Implementation} \label{sec:pe-impl}

SIMPL overloads every operator and special form so that they can handle both syntax objects (Racket's representation of a program) as well as concrete values.  If any of the arguments are syntax, the operator will emit residual code (which is a syntax object) to perform the operation later.  Operations that are by default dynamic (random number generation, mutable data, etc.) will always emit residual code.  PE is then achieved by running the program with these overloaded operators.  Each operator will then build the residual program from the residual programs of its subexpressions.  Note that this implementation automatically takes care of binding-time analysis---the decision to perform an expression depends on whether its subexpressions are concrete, that is, the binding times of the subexpressions.

\lstinline{static} is implemented using a global parameter.  All operations first check this parameter and if static mode is set, then they will attempt to fully evaluate to a concrete value.  Then, for \lstinline{(static expr)}, we set the parameter to static mode, evaluate \lstinline{expr}, and then reset the parameter.

\lstinline{lift} makes a static expression dynamic.  Since a static expression evaluates to a concrete value $c$, \lstinline{lift} makes it dynamic by returning the residual code \lstinline[mathescape]{(quote $c$)}.

\subsubsection{Primitive Procedures} \label{sec:pe-prim-impl}

Consider the following code:

\begin{lstlisting}
(define (square x) (* x x))
(for ([i 10])
  (print (square (+ i 1))))
\end{lstlisting}

Since the loop is not unrolled, the value of \lstinline{i} is unknown, and we must generate residual code for \lstinline{(+ i 1)}.  The na\"{\i}ve approach would be to simply return the expression unchanged.  However, after unrolling the call to \lstinline{square}, the final residual code would be:

\begin{lstlisting}
(print (* (+ i 1) (+ i 1)))
\end{lstlisting}

Note that the \lstinline{(+ i 1)} expression has been duplicated.  This could slow down the program.  The problem becomes worse if we replace \lstinline{(+ i 1)} with an expression with side effects---in this case, by duplicating the code, we have changed the meaning of the program and so the PE would become incorrect.

Thus, in general when partially evaluating an expression, we need two invariants to hold:
\begin{enumerate}
\item The returned value should be either the concrete value of the expression or a piece of code that at runtime will be the value of the expression.
\item All of the subexpressions should be evaluated the same number of times as they would be in the original program.
\end{enumerate}

SIMPL enforces the second invariant by generating a temporary variable for every expression and returning that variable, and so the temporary variable is used where the expression would have been used.  We also add an assignment statement to the residual program binding the temporary variable to the value of the expression.  This forces each expression to be evaluated just once.  This is analogous to SSA form---each temporary variable is written to exactly once.  These assignments are stored in the \emph{program trace}.  For the previous program, SIMPL generates:

\begin{lstlisting}
(for ((i 10))
  (let ()
    (define v1 (+ i 1))
    (define v2 (* v1 v1))
    (print v2)))
\end{lstlisting}

\begin{figure}
\begin{lstlisting}
(define (+^ x y)
; If either argument is residual code
  (if (or (syntax? x) (syntax? y))
; Generate a new temporary variable
      (let ([z (generate-temporary 'v)])
; Bind z to the expression in the trace
        (stmt #`(define #,z (+ #,x #,y)))
; Return the temporary variable z
        z)
; x and y are numbers, so perform the addition
      (+ x y)))
\end{lstlisting}
\caption{Lifting \lstinline{+} for PE.  This ignores details such as simplifying (+ 0 x) and taking in a variable number of arguments.  \lstinline{stmt} adds a new statement to the program trace.}
\label{fig:+}
\end{figure}

So, during PE every primitive pure function performs the following steps:

\begin{enumerate}
\item If all of the arguments are concrete, perform the function application and return the result.
\item Otherwise, generate a new temporary variable v.
\item Add the code \lstinline{(define v expr)} to the \textit{program trace}.
\item Return v.
\end{enumerate}

The code for the lifted \lstinline{+} procedure is shown in Figure~\ref{fig:+}.

Primitive procedures with side effects are handled similarly.  They always generate residual code, except when in static mode.

\subsubsection{Special Forms}

Special forms that do not introduce new variables are handled very similarly to pure functions.  For \lstinline{if}, the expression is performed at PE time as long as the condition is known, even if the two branches are not known.  Other special forms such as \lstinline{cond}, \lstinline{and}, and \lstinline{or} are rewritten in terms of \lstinline{if}.

Special forms that introduce new variables are trickier.  Since we use Racket's evaluation system instead of writing our own evaluator, our variables must be defined in Racket.  There is a neat trick to make this work---during PE, we bind the new variable to code that represents that same variable.  Then, when this variable is used in subsequent expressions, it will evaluate to code that represents the variable itself, which satisfies the required invariants (notably, that evaluating the variable yields residual code that will evaluate to the correct value at runtime).  For example, the lifted \lstinline{for} loop is shown in Figure~\ref{fig:for}.

\begin{figure}
\begin{lstlisting}
(define-syntax (for^ stx)
  (syntax-case stx ()
; Match against a template
    [(for^ ([i itr] ...) expr ...)
; Generate some temporary variables
     (with-syntax ([(v ...)
                    (generate-temporaries
                     #'(i ...))])
      (syntax/loc stx
; Evaluate the iterables, store in vs
         (let* ([v itr] ...)
          (let ([body
; Partially evaluate
                  (pe (thunk
; With the loop variables bound to residual
; code that represents those variables
                      (let ([i #'i] ...)
; The original body
                         expr ...)))])
; Generate the new for loop
             (stmt #`(for ([i #,v] ...) 
                      #,body))))))]))
\end{lstlisting}
\caption{Lifting \lstinline{for} loops for PE.}
\label{fig:for}
\end{figure}

\section{Caching and Code Generation}

The other optimizations are caching and code generation.  Due to lack of space, we do not describe them in detail.

\subsection{Caching}

The \lstinline{cache} construct automatically creates a cache for the argument expression.  \lstinline{cache} dynamically determines which values affect the result of the expression.  In the likelihood weighting for the Burglary model, \lstinline{cache} determines that the weight depends only on $A$, leading to a small, compact cache (see Figure~\ref{fig:likelihood-gen}).

As with most memoization constructs, an incorrect use of \lstinline{cache} can change the behavior of the program.  To facilitate debugging, we provide a debug mode for SIMPL, in which the cache will be created and updated like normal, but when there is a cache hit, the original computation will be run anyway to check whether it agrees with the cache hit.

\subsection{Code Generation}

As can be seen from Figure~\ref{fig:likelihood-gen}, the output of the PE stage is very similar to assembly, but in Racket.  As a result, we only generate code for a small subset of Racket that includes this assembly-like Racket.  For this small subset, code generation to C is very simple---a simple recursive descent over the program is sufficient.  However, if an inference algorithm uses extra unsupported Racket features, then the code generator will fail to compile the code and will signal an error.

\section{Model Data Structure} \label{sec:model}


Since we expect many inference algorithms to be written for a single model data structure, it makes sense to aggressively optimize the model.  In addition, we have found data structure choice crucial for generating fast code.

SIMPL uses a more complex model data structure than the Bayes Net data structure presented in Section \ref{sec:overview}.  In particular, nodes in the model can have arrays of values instead of just a single value.  This allows SIMPL to scale up to solve large problems with lots of evidence.  However, SIMPL does not support all models that PPLs like BLOG or Church can support yet---this is planned for future work.

\subsection{Self-Specialization}

The main optimization in the model is to provide special, more optimized code generation.  For example, in the likelihood weighting example, in the generated code the model was completely eliminated and replaced by boolean-valued variables.  This is not done by PE by itself---the partial evaluator would have kept the model structure and the calls to \lstinline{value}.  The model itself knows that for an algorithm that at runtime only needs to know the values of nodes, it is better to replace the model with boolean variables.  We can consider this to be a \textit{self-specializing data structure}, that is, a model data structure that knows how to specialize itself based on what operations will be needed at runtime.

The implementation of self-specialization is straightforward, albeit a bit tedious.  Instead of just being a data structure, the model consists of a data structure as well as code generating procedures that can specialize the model under various conditions.  For example, when we only need to know the values of nodes at runtime, the model's code generator will generate boolean variables instead of graph nodes.

\section{Evaluation}

\begin{table}[]
\centering
\begin{tabular}{ll}
\toprule
Name & Description \\
\midrule
Burglary & Example in Figure~\ref{fig:burglary} \\
\midrule
CSI & Toy model (4 nodes, 3 edges) with \\
& context-specific independence \\
\midrule
MultiBurglary & 1,000 potential Burglaries and \\
& Alarms, 1 potential Earthquake \\
\end{tabular}
\caption{Benchmark Models for SIMPL.}
\label{tab:models}
\end{table}

\begin{table}[]
\centering
\begin{tabular}{lccc}
\toprule
Name & Programmer Time & LoC & \# Annotations \\
\midrule
Gibbs & N/A & 23 & 3 \\
\midrule
MH & 3 hours & 23 & 3 \\
\midrule
Likelihood & 2 hours & 15 & 2 \\
\midrule
Rejection & 2 hours & 18 & 2 \\
\end{tabular}
\caption{Effort put in to writing inference algorithms.  Gibbs sampling does not have programmer time data because it was developed in tandem with SIMPL over a long time.}
\label{tab:algorithms}
\end{table}

We evaluate the various optimizations in SIMPL by comparing their performance to baseline systems in C and Racket on multiple models and inference algorithms.  We consider three models (Table~\ref{tab:models}) and four inference algorithms (Table~\ref{tab:algorithms}).  We also evaluate the benefits of model specialization.

All experiments were conducted on a Lenovo Ideapad U510 with in i7 core processor.  Every experiment was repeated 10 times and the results were averaged.  For every experiment, the difference between the minimum and maximum values was within 10\% of the average, which is negligible compared to the size of the results, and so error bars are not shown.

Every experiment is designed so that the generated code will be taking samples for at least 0.5 seconds, so that we can be confident in the time measurements.  All experiments measure only the time taken to run the generated code.  In particular, Racket start-up time, model construction time, and code generation time are not included.  However, all of these take negligible amounts of time (each under a second).

\subsection{Contribution of each Optimization}

\begin{figure}
\centering
\includegraphics[scale=0.65]{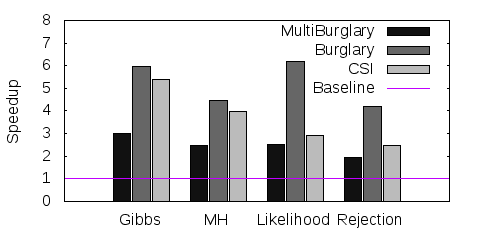}
\caption{Speedups obtained when running SIMPL with only PE, relative to the Racket baseline.}
\label{fig:pe-performance}
\end{figure}

\begin{figure}
\centering
\includegraphics[scale=0.65]{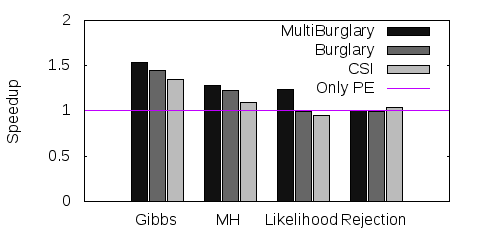}
\caption{Speedups obtained when running SIMPL with PE and caching, relative to only PE.}
\label{fig:cache-performance}
\end{figure}

\begin{figure}
\centering
\includegraphics[scale=0.65]{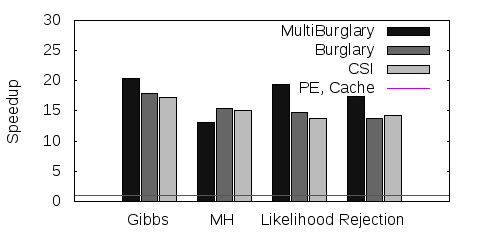}
\caption{Speedups obtained from code generation with SIMPL, relative to only PE and caching.}
\label{fig:c-performance}
\end{figure}

We evaluate the three main optimizations in SIMPL (PE, caching and code generation) and determine the level of speedup we get from each one.  We use four inference algorithms (Rejection sampling, Likelihood weighting, Metropolis-Hastings, and Gibbs sampling), as well as three models and queries (MultiBurglary, Burglary and CSI).

The MultiBurglary example has a House and Alarm node, each of which contain a vector of 1,000 values representing the various houses and alarms.  Thus, the structure of the model is small (containing only three nodes), but it works with a large amount of data.

The Racket baseline system is a library that exports all identifiers that SIMPL exports, but without PE or caching.  Thus, algorithms in SIMPL can automatically be run in vanilla Racket simply by changing the \#lang declaration of the program.  The annotations become no-ops, and a comparison against a completely unannotated program shows this adds less than 2\% overhead.

We evaluate SIMPL in three modes.  In the first mode, SIMPL performs only PE and generates Racket code.  Note that, in this mode, any operation in the generated code was present in the original code---the only change is that some operations have been already performed.  We add caching in the second mode but still generate Racket code.  In the final mode, we perform PE and caching, and generate C code.

The speedup from PE is shown in Figure~\ref{fig:pe-performance}.  The speedup from \textit{adding} caching is shown in Figure~\ref{fig:cache-performance} (this is measured relative not to the baseline, but to SIMPL with only PE).  Similarly, Figure~\ref{fig:c-performance} shows the additional speedup of generating C code.  We can see that PE provides a 2-6x speedup, caching gets an additional 1-1.5x, and generating C code yields 13-20x, for an overall speedup of 30-150x.

Sometimes, caching can be harmful, such as for the CSI model with likelihood weighting.  This is because in some cases, the time taken to look in the cache is larger than the time taken to recompute the expression.  This could be solved by giving a model-specific policy which says when a cache should be used depending on the properties of the model, but we leave this to future work.

\subsection{Comparison to C}

\begin{figure}
\centering
\includegraphics[scale=0.65]{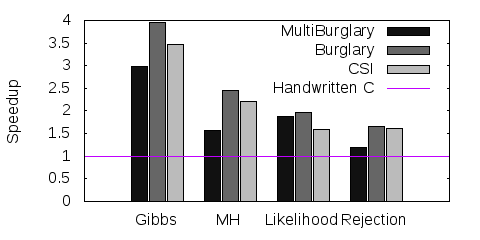}
\caption{Speedups obtained from SIMPL, relative to a baseline C implementation.}
\label{fig:c-baseline}
\end{figure}

The C baseline system is a handcoded implementation of Bayes Nets with the same structure as the Racket system.  In particular, it supports exactly the same models and algorithms as SIMPL, with a similar model data structure, but does not do PE or caching.  So, in this system, the algorithms are \textit{not} specialized.  Figure~\ref{fig:c-baseline} compares SIMPL with all optimizations to this baseline system.  The algorithms are arranged in approximately descending order of complexity and accuracy.  Notice that the more complex and accurate algorithms are also the ones that get the most speedup.  Thus we would expect that new, more complicated inference algorithms would also benefit greatly from SIMPL.

\subsection{Model Specialization}

Model specialization saves us from the performance penalty of a more general model data structure.  We consider two model data structures---the simple Bayes Net model of section \ref{sec:overview}, where each node has exactly one boolean value, and the more general model structure used by SIMPL, where nodes can have arrays of values.  The Burglary and CSI models can be expressed in both cases, but MultiBurglary can only be expressed in the more general model structure.  For Burglary and CSI, due to the specialization in the general model structure, the code generated for the model is identical for both the general and the simpler models, and we do not pay a penalty for the generality at runtime.

\begin{figure}
\centering
\includegraphics[scale=0.65]{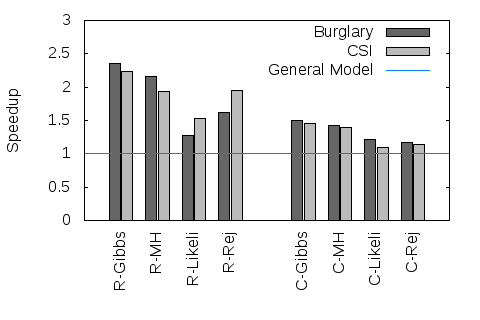}
\caption{Speedups in baseline implementations from switching to simpler models.  Left - Racket, Right - C.}
\label{fig:model-graph}
\end{figure}

For comparison, we used a simpler model in the two baseline systems, and measured the speedup obtained by switching to the simpler model for Burglary and CSI.  The results are shown in Figure~\ref{fig:model-graph}.  There is a large speedup for Racket and a significant but smaller speedup for C.  These speedups are for a small difference in model complexity---as we generalize to all models supported by PPLs, we expect huge slowdowns in baseline systems.  Due to model specialization, we would expect less of a performance penalty for SIMPL.

\subsection{Simplicity}

There is some anecdotal data that suggests that SIMPL is easy to use.  Table~\ref{tab:algorithms} shows that it takes very little time (2-3 hours) to implement a new annotated inference algorithm, and there are only 2-3 annotations for each algorithm.

SIMPL is also robust to changes in the program---for example, if \lstinline{true-cp} is implemented using \lstinline{for/fold} instead of recursion, the baseline slows down by 1.5x, but the generated code is nearly identical.  This is some evidence that SIMPL is more transparent and consistent than traditional compilers---equivalent programs produce equivalent generated code, even though they have different performance in vanilla Racket.

\subsection{Existing Solutions}

SIMPL is 2-3 orders of magnitude faster than the BLOG interpreter, and 2-3x faster than Swift, the BLOG compiler.  However, this is not a fair comparison---Swift supports a much larger set of models compared to SIMPL, and so we expect that it pays a price for generality.  When SIMPL supports the same variety of models, we expect its performance will be similar to that of Swift.  However, a major advantage of SIMPL is that adding a new inference algorithm is easy.  Adding a new inference algorithm to the BLOG compiler is a huge effort---in fact, because of this, the compiler currently supports only two inference algorithms.

We expect that the optimizations in SIMPL will extend to more general models as well.  For example, the graph traversal in inference algorithms can still be partially evaluated away.  Other optimizations such as the optimization for \lstinline{cp} would have analogs in more general models.  For example, the sampling procedure could be specialized if the mean and variance are known.

\section{Related Work}

While some PPLs do have compilers, to our knowledge SIMPL is the first compiler that aims to make it easy to add new inference algorithms.  For example, Bher \cite{Bher} compiles Church, but only supports MCMC sampling.

The R2 system \cite{R2} propagates evidence backwards through a probabilistic program to reject samples as early as possible.  While this has the goal of making inference faster, the optimization is orthogonal to ours---it reduces the number of calls to the random number generator (by removing some sampling steps), whereas we reduce overhead without affecting the number of calls to the random number generator.  In fact, we could first use the R2 optimization to generate a new, better probabilistic program, and then use our system to optimize inference over the new model.  This also holds for other optimizations such as program slicing \cite{Slice}.

PE has been studied extensively, and an overview can be found in \cite{Sestoft}.  One recent approach to automatic PE is multi-stage programming languages such as MetaML \cite{MetaML}.  However, multi-stage programming requires a statically computed binding-time analysis, which precludes operations like \lstinline{(value node)} which are sometimes evaluated and sometimes generate residual code.

Another approach to automatic PE is Lightweight Modular Staging \cite{LMS}.  LMS is similar to our approach in that it is value-driven, and we could probably obtain similar results using it, but this would require more effort from the inference algorithm writer.  SIMPL provides a simpler interface to the programmer through annotations.

The Delite framework \cite{Delite} allows developers to create DSLs and automatically get benefits of common optimizations.  It may be feasible to implement our DSL in Delite, but we do not think the primitive operations provided by Delite (which focus on collections) are well suited for inference.

Programmer controlled compilation has been added to the JIT in Lancet \cite{Lancet}, which exposes an API for the JIT compiler to the programmer, so that the programmer can control compilation.  Our work is similar, but since we target only inference algorithms, we provide higher-level optimizations that are more accessible for an algorithm writer.

Memoization has been extensively studied, and many libraries and techniques exist \cite{Acar} \cite{Knoblock}.  It has even been used to speed up probabilistic inference \cite{Pfeffer}.  In future work, we intend to integrate these techniques with the \lstinline{cache} construct.

Broadway \cite{Broadway} uses annotations to provide semantic information about a library in order to optimize it.  It replaces general library calls with manually written specialized versions, whereas we want to specialize the code automatically.

\section{Conclusion and Future Work}

We have presented SIMPL, a DSL for automatically specializing inference algorithms, that uses PE, caching, and good data structures to tremendously speed up inference algorithms.  SIMPL leverages knowledge about the probabilistic model to generate efficient code that outperforms code produced by the optimizing compilers of Racket and C.

Currently, SIMPL cannot work with as many models as existing PPLs.  We intend to extend SIMPL to work with more types of probability distributions and with open-universe models \cite{OpenUniverse}.  We expect that some optimizations, such as the PE of the model graph traversal, will carry over with minimal change, and that other optimizations will have analogs in more general models.

The \lstinline{cache} annotation in SIMPL performs a sophisticated analysis of model dependences, but is lacking in other areas identified in existing research \cite{Acar}.  We hope to integrate other approaches in future work.

Currently, the annotations for SIMPL are programmer-supplied and model-independent.  We could instead implement an autotuner, as done in Halide \cite{Halide}, or we could allow the programmer to specify a \textit{policy}, in which annotations depend on model properties---for example, we could have a \lstinline{for/unroll/when} annotation that unrolls loops when an input-dependent condition is met.

\bibliographystyle{plain}
\bibliography{references}
\end{document}